\input harvmac.tex

\def\nup#1({Nucl.\ Phys.\ $\bf {B#1}$\ (}
\def\l{\lambda}
\noblackbox
\Title{\vbox{
\hbox{HUTP-98/A070}
\hbox{\tt hep-th/9812127}
}}{M-Theory and Topological Strings--II}
\bigskip
\centerline{Rajesh Gopakumar \foot{gopakumr@tomonaga.harvard.edu}and  Cumrun
Vafa \foot{vafa@string.harvard.edu}}
\bigskip
\centerline{Lyman Laboratory of Physics}
\centerline{Harvard University}
\centerline{Cambridge, MA 02138}

\vskip .3in
It is shown how the topological string amplitudes encode
the BPS structure of wrapped M2 branes in M-theory compactification
on Calabi-Yau threefolds.  This in turn is related to
a twisted supersymmetric index in 5 dimensions which
receives contribution only from BPS states.
The spin dependence of BPS states in 5 dimensions
is captured by the string coupling constant dependence
of topological string amplitudes.

\Date{December 1998}
\lref\naret{I. Antoniadis, E. Gava, K.S. Narain and T.R. Taylor,
``Topological Amplitudes in String Theory'', \nup413 (1994), 162.}
\lref\narcon{I.Antoniadis, E.Gava, K.S.Narain, T.R.Taylor,
``N=2 Type II- Heterotic duality and Higher derivative
F-terms'', \nup455 (1995), 109.}
\lref\bcovi{M.Bershadsky, S.Cecotti, H.Ooguri, C.Vafa , ``Holomorphic
Anomalies in Topological Field Theories, \nup405 (1993), 279 .}
\lref\iz{C. Itzykson, J. Zuber, ``Quantum field theory'', Addison Wesley
Publishing.}
\lref\bcovii{M. Bershadsky, S. Cecotti,H. Ooguri, C. Vafa ,
``Kodaira-Spencer Theory of Gravity and Exact Results for
Quantum String Amplitudes'', Comm. Math. Phys. {\bf 165} (1994) 311 .}
\lref\mtop{R. Gopakumar, C. Vafa, ``M-Theory and Topological Strings -I'',
hep-th/9809187 .}
\newsec{Introduction}
Topological string theories \ref\wittop{E. Witten,
``Topological Sigma Model,'' Comm. Math. Phys. 118 (1988) 411
\semi ``Chern-Simons Gauge Theory As a String Theory,'' hep-th/9207094.}\
are among
the most beautiful and simple string theories.  In particular
in connection with target space being Calabi-Yau 3-folds they
have a rich structure.  Their definition (in the A-model) involves
summing over holomorphic maps from worldsheets to the Calabi-Yau
3-fold.
In this sense, they seem to be related to counting some BPS 2-brane
states, such as M2/D2 branes in M/IIA compactifications
on Calabi-Yau threefold.  A natural question is to find the
precise connection between the BPS content involving
M2/D2 branes and the topological string amplitudes.
One aim of the present paper is to make this connection more precise.
The key ingredient in this connection is the reinterpretation
of topological string amplitudes on Calabi-Yau threefolds as
computing particular F-terms in $N=2$ supersymmetric theories
in 4 dimensions \bcovii \naret.   Our basic strategy is to find the
contribution
of BPS states to F-terms and thus
read off the BPS content of topological string amplitudes.
Some steps in this direction had already been taken in
\ref\cvt{C. Vafa, ``A Stringy Test of the Fate of the Conifold,''
Nucl. Phys. {\bf B447} (1995) 252.}\ref\harmor{J. Harvey and G. Moore,
``Algebras, BPS States, and Strings,''
Nucl.Phys. {\bf B463} (1996) 315.}\ref\kmv{A. Klemm,
P. Mayr and C. Vafa, ``BPS States of Exceptional Non-critical Strings,''
Proc. of the conf. {\it Advanced Quantum Field Theory}, (in memory
of Claude Itzykson), hep-th/9607139.}.

This paper is a continuation of our previous paper
\mtop\ where we considered the contribution
of the simplest BPS states to the topological string amplitudes.
In that paper it was useful to view the D2 brane BPS content of Type
IIA on Calabi-Yau 3-fold, from M-theory perspective compactified
on the same Calabi-Yau times a circle.  The momentum
modes around the circle were crucial in reading off from
the M2 brane spectrum, the structure of D2/D0 brane contribution.  Similarly
in this paper, we find that the M-theory perspective is useful.  In
fact this link is even more important in connection with the present
paper:  It turns out that off shell quantum field content
of BPS states in type IIA string is necessary in order to evaluate
their contribution to topological string amplitudes.  On the other
hand the D2 brane bound state quantum numbers do not yield
the corresponding off shell quantum field.  Whereas, the M2 brane
{\it on shell} state, will give rise to the corresponding
{\it off shell} field in 4 dimensions.  It is this
link between M-theory and type IIA BPS states which proves
crucial for connecting topological string amplitudes to BPS structure
of M-theory on CY threefold.  Along the way we introduce
a generalized supersymmetric index in 5 dimensions, somewhat
analogous to similar objects introduced in 2 dimensions \ref\nsin{S. Cecotti,
P. Fendley, K. Intriligator and C. Vafa,``A New Supersymmetric Index,''
Nucl. Phys. {\bf B386} (1992) 405.},
which captures the BPS content of the theory.

The organization of the paper is as follows:  In section 2 we
show how to compute the contribution of a single BPS state
in 4 dimensional compactification of type IIA strings to topological
string amplitudes.  This turns out to be a straightforward
one loop computation generalizing Schwinger's computation
to this setup.  In section 3 we connect the M-theory
content of BPS states with those of type IIA and use
that to rewrite the topological string amplitudes in terms of
BPS content of M-theory in 5 dimensions. This section contains
the main result of this paper.
 In section 4 we show how in some cases one can compute the BPS spectrum
explicitly, by studying M2/D2 branes wrapped over a genus $g$
Riemann surface in Calabi-Yau.

\newsec{Schwinger Re-interpretation of Topological String Amplitudes}

Topological A-model string amplitude at genus $g$
roughly speaking computes the weighted sum over holomorphic maps
from a Riemann surface of genus $g$ to a Calabi-Yau threefold:
$$F_g(t_i) =\sum_{{\rm Hol. Maps.}}{\rm exp}(-2\pi A)$$
where $A=\sum n_it_i$ denotes the area of the surface in
the Calabi-Yau $M$ relative to a (complexified)
Kahler form parameterized by $t_i$
in terms of a choice of $H_2(M)$ and $n_i$ denotes
the $H_2$ class of the image.   The sum over the holomorphic
curves
in the above is sometimes replaced by an appropriate
integral when there are families of holomorphic curves.
The topological string amplitudes compute
corrections of the form $F_g(t_i)R_+^2 F_+^{2g-2}$ in type IIA
compactifications on a Calabi-Yau \bcovii \naret , where $R_+$ and
$F_+$ denote the self-dual parts of Riemann and graviphoton
field strengths\foot{This is strictly speaking valid for $g>0$, however
our expressions continue to be valid also for $g=0$ when they
are appropriately interpreted as computing prepotential terms in the
corresponding $N=2$ theory.}.
It is natural to consider giving vev to $F_+=F$ and denoting
$g_sF=\lambda$.  Then the topological string partition function
defined as
$$F(\lambda ,t_i)=\sum_{g}\lambda^{2g-2}F_g(t_i)$$
can be viewed as computing the correction of the
form $F(\lambda ,t_i) R_+^2$ in type IIA compactification
on a Calabi-Yau threefold.

It is well known that these amplitudes do not receive
any further perturbative or even non-perturbative string correction beyond
genus $g$
(due to the decoupling of the dilaton, which is in a hypermultiplet, from
kahler moduli which are in vector multiplets).  Thus one can evaluate
them at strong coupling, in which case one would expect the dominant
contribution to come from light D-brane bound states.  Moreover,
due to the fact that it is an F-type term, one expects this
to receive contribution only from BPS states involving D-branes.
In the limit of a large volume Calabi-Yau, the relevant D-brane states would be
D2 branes wrapped around cycles, bound to D0 branes.  The effect
of integrating these fields out should be to reproduce the above
amplitude.  Given the minimal couplings of these fields to supergravity
multiplets, one is thus reduced to doing a one loop computation with
bound states of D-branes going around the loop with emission
of graviphoton and graviton fields.

It is crucial to
note that the contribution of the fields
at one loop {\it will depend} on their off-shell quantum fields (see
e.g. \ref\duff{S.M. Christensen and M.J. Duff, Nucl. Phys. {\bf
B 154} (1979)340.}).
In particular knowing the $SO(4)=SU(2)_L\times SU(2)_R$ Lorentz group
content of the fields
is necessary.  An $N=2$ BPS state is represented by a field content of the
form
$$[({1\over 2},0)\oplus 2(0,0)]\otimes \sum_{j_1,j_2}N_{j_1,j_2}(j_1,j_2)$$
for some integers $N_{j_1,j_2}$.  Before computing
the contribution of such states, let us note what
it could depend on. First of all, we can compute the contribution
for each $(j_1,j_2)$ and add the total result weighted with $N_{j_1,j_2}$
to get the final result.
 In addition to its mass (or more precisely the
BPS central term $Z$), the $SU(2)_R$ content of the fields should be
irrelevant for the contribution to $R_+^2 F_+^{2g-2}$ amplitudes.
The reason for this is that $R_+$ and $F_+$ by definition only
couple to the $SU(2)_L$ quantum number.  In other words, only
the degeneracy and fermi/bose statistics is relevant in computing
the contribution of $SU(2)_R$ quantum numbers. Thus,
the relevant contribution is $\sum_{j_2} (-1)^{2j_2}(2j_2+1) N_{j_1,j_2}$
times the contribution of the state with left-spin
$[(\half )+2(0)]\otimes [j_1]$.

As in \mtop , the Schwinger one-loop calculation for a particle in a background
constant electromagnetic field, will be a surprisingly powerful way of
determining the structure of higher genus topological partition functions
$F_g$.  This is natural in view of the fact noted above that the
topological string partition function arises naturally by considering
constant self-dual field strength for the graviphoton field.
Let us first recall Schwinger's computation. Schwinger (see for example \iz )
studied the
case of a particle minimally coupled to a constant background
electromagnetic field. Using the proper time formalism,
the resulting determinant (a one loop contribution)
can be explicitly evaluated. For instance, the
free energy of a charged scalar (of mass $m$)
in a constant self-dual field of magnitude
$F$ is
\eqn\sch{{\cal F}=\int_{\epsilon}^{\infty}
{ds\over s}{\rm Tr}e^{-s( \triangle +m^2)}
={1\over 4}\int_{\epsilon}^{\infty}
{ds\over s}{1\over {\rm sinh}^2{seF\over 2}}e^{-sm^2}.}
Note that ${\cal F}$ can be expanded in powers of ${eF\over m^2}$ -- the
only dimensionless combination. These terms are the quantum corrections to
Maxwell's action due to the presence of charged particles. There are in
addition, non-perturbative pieces which go like ${\rm
exp}{-{m^2\over  eF}}$.

When the charged particle carries non-trivial spin, then the answer is
modified to
\eqn\schspin{{\cal F}=\int_{\epsilon}^{\infty}
{ds\over s}{\rm Tr}(-1)^Fe^{-s(\triangle+m^2+2eJ\cdot F)}
={1\over 4}\int_{\epsilon}^{\infty}
{ds\over s}{1\over {\rm sinh}^2{seF\over 2}}e^{-sm^2}
{\rm Tr}(-1)^F e^{-2se J\cdot F}.}
where $J=J_R+J_L$ is the generator of spin angular momentum $J\cdot F
\equiv J_{\mu \nu}F^{\mu \nu}$.

What does all this have to do
with topological string amplitudes? As we mentioned,
we are really computing the $R_+^2 F_+^{2g-2}$ terms in the four
dimensional effective action. As argued earlier, these should be given by a
one loop computation where we integrate out D0/D2 branes. This seems like a
generalization of the Schwinger computation to the case where one has both
background electromagnetic and metric fields. In other words, we want the
$R^2$ term in a heat kernel expansion in this background. This can be
quite tedious. Fortunately, the answer turns out to be simple due to
supersymmetry.  Basically, what we find is that the
effect of $N=2$ supersymmetry is to convert the Schwinger's
computation for the vacuum amplitude in the presence of constant
electromagnetic field strength in the non-supersymmetric case, to the
correction for $R^2$ in the supersymmetric case.

For instance, in the case of an isolated $S^2$,
lightest states at strong coupling,
other than the 0-branes, are wrapped 2-branes.
The 0-branes as well as the $S^2$
wrapped 2-brane are BPS states
in the $[({1\over 2},0)\oplus 2(0,0)]$ representation of the
Lorentz group $SO(4)=SU(2)\times SU(2)$.
The contribution of such a hypermultiplet to $F_gR_+^2 F_+^{2g-2}$
is exactly that of a charged {\it scalar} to $F_+^{2g-2}$ in the
ordinary Schwinger computation \narcon . Roughly speaking, the fermionic
contribution compensates for the extra insertions of the curvature.

Thus the contribution of such a hypermultiplet to ${\cal F}_g$ is that in
\sch\ (note that here $|Z|=|m|=e$, where $Z$ is the central term
in the $ N=2$ algebra)
\eqn\schw{\sum_{g=0}(Fg_s)^{2g-2}{\cal F}_g={1\over 4}\int_{\epsilon}^{\infty}
{ds\over s}{1\over {\rm sinh}^2{sF\over 2}}e^{-s{Z\over g_s}}.}
which gives
\eqn\conh{{\cal F}_g= -\chi_g Z^{2-2g}}
where $\chi_g=\chi({\cal M}_g)$ is the euler characteristic of the moduli
space of genus $g$ Riemann surfaces.

Though this was evaluated in a heterotic computation in \narcon\
the final answer was extracted in a field theory
limit. This will help us to get the contribution of a general N=2 BPS state.
Since the stringy one loop computation had particles of arbitrary spin
inside the loop, we can also extract the respective contributions
to $R_+^2F_+^{2g-2}$.
We might guess from our experience with the
hypermultiplet that the contribution
to $R_+^2F_+^{2g-2}$ from a state with spin
$[({1\over 2},0)\oplus 2(0,0)]\otimes [(j_1,j_2)]$
is the same as that to $F_+^{2g-2}$ from a particle of spin
$[(j_1,j_2)]$ from a Schwinger computation. This will prove to be right.

Even though one can do a Schwinger computation at one loop in
a field theory setup,
it turns out that the one loop heterotic setup is more convenient.
As already mentioned, this computation has already been done in \narcon .
The fermionic zero modes at one loop absorb the $R^2$ fermion
vertex fields and the effect of having a constant self
dual graviphoton field strength is to modify the spacetime
contribution to quadratic
bosonic
path integral on a torus with modulus $\tau$ (Eq. 4.13 in \narcon )
\eqn\narpath{\eqalign{G(F, \tau, \bar{\tau})=
&{\int\prod_{i=1,2}DZ^iiD\bar{Z}^i
exp(-S +\int{F\over \tau_2}(Z^1\bar{\del}Z^2
+\bar{Z}^2\bar{\del}\bar{Z}^1)d^2\sigma)\over \int\prod_{i=1,2}DZ^iiD\bar{Z}^i
exp(-S)} \cr =& ({2\pi iF\bar{\eta}^3\over
\bar{\Theta}_1(F,\bar{\tau})})^2 \exp(-{\pi F^2\over \tau_2})}}
Here $Z^i$'s are complex bosonic fields representing spacetime variables
with the usual free field action
denoted by
$S$. And $F$ as before, is the expectation value of the background self-dual
electromagnetic field. $\Theta_1$ is the Jacobi theta function.

This formula has a straightforward Hamiltonian interpretation which can be
read off from the path integral in \narpath . The $\l$ dependent term
is that from a self-dual electromagnetic field in the four dimensions
labelled by $Z^1, Z^2$. This can also be seen from the answer in the
second line of Eq.\narpath\ which can be written as
\eqn\narg{G(F, \tau, \bar{\tau})=F({\pi F\over \sin(\pi F)})^2
{\prod_{n=1}(1-q^n)^4\over \prod_{n=1}(1-e^{2\pi i F}q^n)^2
\prod_{n=1}(1-e^{-2\pi i F}q^n)^2}\exp(-{\pi F^2\over \tau_2})}
(In
comparing with the conventions we are using we should make the replacement
$F\rightarrow {F\over 2\pi i}$.)
In \narcon\ only the first term was extracted as the
contribution of a hypermultiplet which we saw in Eqn\schw\ .
But now it is easy to read off the contribution of
higher spins.  Note that the spin content of an elementary
BPS state in the heterotic string comes from ground state
oscillators on the right-moving sector (the supersymmetric side)
tensored with a tower of left-moving bosonic oscillators. In particular
the spacetime quantum numbers of fields will come from the
left-moving bosonic oscillators.
These consist of four bosonic oscillators $\alpha_{-n}^\mu$ each
of which transforms according to $[\half,\half]$ representation
of $SO(4)$.  Given the fact that we have turned on only
$F=F_+$, it is easy to see that the
the denominator in the second term
indicates the $\rm{Tr}\exp(-\tau L_0+2J_3^L F_+)$ structure for
the bosonic string oscillators.

So the contribution to the $R_+^2$ term from a general particle is given by
the generalization of Eqn.\schw\
\eqn\schwgen{\sum_{g=0}R_+^2(g_s F)^{2g-2}{\cal F}_g =(-1)^{2j}R_+^2{1\over 4}
\int_{\epsilon}^{\infty}{ds\over s}
{\rm{tr}e^{-2sJ^L_3 F}\over {\rm sinh}^2{sF\over 2}}
e^{-s{Z\over g_s}}.}
where $j=j_L+j_R$.
Since $g_sF$ always appears in a single combination, we will henceforth denote
the
product by $\l$. We will also rescale $s$ in all the Schwinger integrands
to write it solely in terms of $\l$.

It is simple to use this to
evaluate the contribution from a particle of a given spin. One just carries
out the sum over the helicities signified by the trace.
As already noted, only the left spin content of
the multiplet is relevant for this contribution.
It will turn out to be convenient to write down the contribution from the
supersymmetric multiplet whose left spin content is given by
$I_1\otimes I_r \equiv [(\half )+2(0)]\otimes
[({1\over 2})\oplus 2(0)]^r$. The
trace
over the helicities comes out simply to be $(-4)^r {\rm sinh}^{2r}{s\over 2}$
(For $r=1$
one has for example  $(2-e^{-2s}-e^{2s})$.
Thus the contribution of $I_r$ is given by
\eqn\hgcont{\sum_{g=0}^{\infty}\l^{2g-2}{\cal F}_g(Z)=
\int_{\epsilon}^{\infty}
{ds\over s}(2i{\rm sinh}{s\over 2})^{2r-2}
{\rm exp}(-s{Z\over \l}).}
{}From this we see that $I_r$ contributes only to the topological partition
function $F_g$ at genus $g\geq r$. Moreover, the $I_r$ have the
utility of serving as a convenient basis for decomposing any supersymmetric
multiplet -- a multiplet with maximum spin $j$ can be written as
$\sum_{r=0}^{r=2j}\alpha_rI_r$ for some (possibly negative)
 integers $\alpha_r$.  (Also note that
$I_r\otimes I_s=I_{r+s}$.)

The contribution to the topological partition function from such a
multiplet then takes the form
\eqn\jmult{\sum_{g=0}^{\infty}\l^{2g-2}{\cal F}_g(Z)=
\sum_{r=0}^{r=2j}\alpha_r
\int_{\epsilon}^{\infty}
{ds\over s}(2i{\rm sinh}{s\over 2})^{2r-2}{\rm exp}(-s{Z\over \l}).}

\newsec{5 dimensional interpretation}
In the previous section we have seen how a BPS state in type IIA
compactification on CY 3-fold
contributes to topological string amplitudes.  We also saw
that the off shell $SO(4)$ Lorentz quantum numbers of the
field describing the BPS states is
necessary to find its contribution.  This assignment
of off shell states, which sounds unnatural for four dimensional
Hilbert space, is quite natural from the perspective of M-theory
Hilbert space.
Recall that type IIA compactified on a CY 3-fold at strong coupling
can be viewed as M-theory compactified on CY 3-fold times a circle $S^1$.
The bound states of D2 branes and D0 branes can be viewed in this set up
as M2 branes wrapped around cycles of Calabi-Yau threefold, together with a
momentum around $S^1$. Thus the spectrum of relevant BPS states
should come, from the M-theory perspective, by simply studying
the BPS spectrum of M2 branes wrapped around cycles of
Calabi-Yau threefold.  Note that the rotation group in
five dimensions is $SO(4)$.  Thus the BPS states will have
quantum numbers in terms of it.  This will also yield, upon
compactification on $S^1$ the off-shell field content of the
corresponding 4-dimensional state.

Let us then consider M-theory compactification
on Calabi-Yau threefold.  Consider
$$I(\alpha , \beta )={\rm Tr} (-1)^{2j_L+2j_R} {\rm exp}(-\alpha J_3^L-\beta H
)$$
where the trace is over all massive one particle states at zero center of mass
momentum.  The supercharges of $N=2$ theories in $d=5$ decompose
under $SO(4)$ spatial rotation group
according to $2({1\over 2},0)+2(0,\half )$.  The Hilbert
space forms various representations according to the action of supercharges:
the long multiplets and the short multiplets which
are the BPS states.  The long multiplets are not killed
by any of the supercharges, whereas the BPS multiplets are killed
either by the $2(\half,0)$ supercharges or $2(0,\half )$.
Thus the long multiplets do not contribute
to the above trace due to the $(-1)^{2j_R}$ term.  Only
the BPS states which are annihilated by $2(0,\half)$ survive
in the above trace.  In this sense this is a ``supersymmetric index''
which receive contribution only from BPS states, in the same
sense as ${\rm Tr}(-1)^F F {\rm exp}(-\beta H)$
considered in \nsin\  in connection with
BPS states of $N=2$
supersymmetric theories in 2 dimensions.  Note that this
index is in particular independent of deformation of complex structure
of Calabi-Yau threefold.  For a given BPS state the masses depend
only on the Kahler class of the Calabi-Yau metric, and so the
only dependence could come from jumps.  In fact
it is known that when one changes the
complex structure of Calabi-Yau there could be new BPS states
(in other words the spectrum of bound M2 branes changes).  However
the change must be such that the new states created or destroyed under
this operation collect themselves into long multiplets that thus
do not contribute to the above index.  In other words, the spectrum
of BPS states by itself will change as one changes the complex
structure of the Calabi-Yau, without changing the above index.
In this sense, this is the more natural object to hope to be
able to compute.  Note that the existence of the above index
is related to the fact that (only) in 5 dimensions rotating black
hole configurations which preserve supersymmetry can exist
\ref\robl{J.C. Breckenridge, R.C. Myers, A.W. Peet and C. Vafa,
``D-branes and Spinning Black Holes,'' Phys. Lett. {\bf B391}
(1997) 93.}.

In application to topological string it is convenient
to characterize the contribution of BPS states to the above index
as follows:
Consider the BPS states with
a fixed central charge, $A=\sum_{i=1}^{H_2(M)} n_it_i$ where
$n_i$ denotes the wrapping class of an M2 brane in $H_2(M)$ relative to a basis
and $t_i$ denotes the integral of the Kahler class on the corresponding
class.  The mass of the corresponding wrapped M2 brane (in our normalization)
is $M=2\pi A$.  Consider the totality of all BPS states
with central charge $A$.  Note that if $(n_i)$ are not relatively
prime this also includes BPS states coming from multiwrapped M2 branes.
We arrange the totality of such BPS states in class
$(n_i)$ according to their $SO(4)$ spin
content:
$$[(\half ,0)\oplus 2(0,0)]\otimes \sum_{j_1,j_2}
N^{(n_i)}_{j_1,j_2}(j_1,j_2)$$
Their relevant contribution to the above index is captured by just
keeping track of the total degeneracy of the right rotation group
while keeping the representation for the left rotation group:
\eqn\bpsf{\sum_{j_1,j_2}(-1)^{2j_2}(2j_2+1) N^{(n_i)}_{j_1,j_2} [j_1]=
\sum \alpha^{(n_i)}_r I_r}
where we have written the left-spin content
in terms of $I_r=([\half]+2[0])^r$, which is a
convenient basis.  In this way of writing $\alpha^{(n_i)}_r$ may be
positive or negative integers.
  The quantity which will appear
in the topological string amplitudes are the integers
$\alpha^{(n_i)}_r$ for each class $(n_i)$ and each integer
$r\geq 0$.

In going down from 5 to 4 dimensions on $S^1$
each BPS state can have in addition an arbitrary momentum $m$
around the circle.  In particular each of them lead
to a BPS state in four dimensions with central charge
$Z=2\pi (A+i m)$ where $A=\sum n_it_i$.  Using the contribution
of a single BPS state in 4 dimensions with central charge $Z$
to the topological string amplitude \jmult , and doing a redefinition
of topological string coupling constant $\lambda\rightarrow  i\lambda$
to make it consistent with the conventional topological definitions,
we see that we can
rewrite the full topological string amplitude as
$$F(\lambda )=\sum_{{(n_i)},r,m} \alpha^{(n_i)}_r \int_\epsilon^{\infty}
 {ds\over
s}(2{\rm sin
{s\over 2}})^{2r-2}{\rm exp}[-2\pi {s\over \lambda}(\sum n_it_i +im)]$$
Rewriting the sum over $m$ using the identity
$$\sum_m exp(-2\pi im{s\over \lambda})=\sum_k \delta({s\over \lambda}-k)$$
allows one to do the integral over $s$ and we obtain
our final formula for the topological string amplitude
\eqn\finfor{F(\lambda )=\sum_{{(n_i)},r\geq 0,k>0} \alpha^{(n_i)}_r {1\over
k}(2{\rm
sin}{k\lambda \over 2})^{2r-2}exp[-2\pi k(\sum n_it_i )]}
This formula encodes all the integrality properties of $F(\lambda )$
in terms of $\alpha^{(n_i)}_r$ which are integers. Moreover it suggests
that we should view these numbers as the fundamental invariants that
topological string captures.  It has been known that using mirror symmetry
one can essentially compute topological string amplitudes
\bcovii .  However for every genus $g$ there are a finite
number of undetermined coefficients which need to be fixed first.
The integrality properties of \finfor\
is a powerful constraint in determining these coefficients
and completely computing
topological string amplitudes,
as has been recently demonstrated in \ref\klemm{A. Klemm, work to appear.}.
 We will now
explore some of the properties of the expression \finfor\ for the topological
string partition function.

We first wish to comment on the usefulness of the basis we have chosen to
write the BPS states \bpsf .  The first fact is that
$\alpha^{(n_i)}_r$ do not contribute to $F_g$ for $g<r$.  In other
words, up to a fixed genus, only a finite
number of $r$'s contribute.  The next point is that the contribution
of $\alpha^{(n_i)}_r$ to $F_g$ for $g\geq r$ is of the form
\eqn\multc{\alpha^{(n_i)}_r C_{r,g} \sum_k k^{2g-3} {\rm exp}(-2\pi k A)}
where $A=\sum n_it_i$ and $C_{r,g}$ are some universal rational numbers
(involving Bernoulli numbers).  Moreover, $C_{g,g}=1$.
This implies that if we have computed topological string
amplitudes up to a given genus $g$, we can inductively extract
all $\alpha^{(n_i)}_r$ for $r\leq g$.   In particular
we can
subtract out the contribution of lower $g$ topological
string amplitudes $F_r$ with $r< g$ from $F_{g}$ and extract
 $\alpha^{(n_i)}_g$ from $F_g$.  This encodes
all the bubbling and multicovering structure that one should
anticipate for topological string amplitudes.  The connection
between multi-covering for genus zero topological
string amplitude and its M-theory
interpretation was already pointed out in \ref\nekla{A. Lawrence and N.
Nekrasov,
``Instanton sums and five-dimensional gauge theories,''
Nucl. Phys. {\bf B513} (1998) 239.}.
The formula \multc\ applies to contribution of non-zero
$(n_i)$.  The case with $(n_i)=0$ has already been considered
in \mtop\ref\mor{M. Marino and G. Moore, ``Counting higher genus curves in a
Calabi-Yau manifold,'' hep-th/9808131.}\ref\fabpand{C. Faber and R.
Pandharipande, ``Hodge integrals and
Gromov-Witten theory,'' math.AG/9810173.}.

\newsec{M2 branes wrapped around genus g surfaces in CY}
We have seen that topological string amplitudes can be
rewritten in terms of BPS degeneracies of M2 branes wrapped
around Calabi-Yau 2-cycles.   In particular the numbers
$\alpha^{(n_i)}_r$, which can be used
to rewrite the topological string amplitudes
according to \finfor , capture the $SU(2)_L$ content of M2
BPS states in the $H_2$ class given by $n_i$.
In this section we would like to study how in principle
one would compute $\alpha^{(n_i)}_r$ by considering
M2 branes wrapped around a genus
$g$ curve in the Calabi-Yau 3-fold.  It is also convenient
to consider the related problem of studying the
D2 brane
bound states.

The worldvolume of a single M2 brane consists of $8$ scalars
transforming as a vector representation of $SO(8)$ R-symmetry group,
and 8 fermions, transforming as $[s\otimes 8_s]$ where $s$ denotes
the 3-dimensional spinor, and $8_s$ denotes the spinor of $SO(8)$
R-symmetry group.  The $SO(8)$ group is the rotation
symmetry normal to the M2 brane worldvolume.  Consider wrapping
the M2 brane around a genus $g$ curve $\Sigma^g$ in a Calabi-Yau 3-fold.
Then the $SO(8)$ is naturally decomposed as
\eqn\deco{SO_J(4)\times SO_N(4) \subset SO(8)}
where we can identify $SO_J(4)$ as the rotation subgroup
of the Lorentz group in 5 dimensional spacetime, and $SO_N(4)$
as the rotation group normal to the M2 brane in the Calabi-Yau threefold.
We can decompose the
$SO(8)$ spinor according to the breakup \deco\
\eqn\spind{8_s\rightarrow [(\half ,0)_J\otimes (\half ,0)_N]\oplus [(0,\half
)_J\otimes
(0,\half )_N]}
where for each $SO(4)=SU(2)^L\times SU(2)^R$ group we have used $(p,q)$ to
denote
the corresponding $SU(2)$ representations.
Being inside a Kahler manifold means that the $SO(4)_N$ holonomy
resides in a $U(2)$ subgroup, $ U(1)_L\times SU(2)_R
\subset SU(2)_N^L\times SU(2)_N^R$.
The tangent bundle of Calabi-Yau, splits over $\Sigma^g$ to
\eqn\scs{T_{CY}\rightarrow T_{\Sigma^g}\oplus N}
where $N$ is a $U(2)$ bundle over $\Sigma^g$.  Moreover, the fact
that the first Chern-class of Calabi-Yau is zero, means that
the determinant bundle $U(1)_L\subset U(2)$ is the same
as $T^*_{\Sigma^g}$, the cotangent bundle on $\Sigma ^g$.

The fields on the worldvolume of M2 brane are twisted
because of the non-trivial embedding in the Calabi-Yau \ref\bsv{M.
Bershadsky, V. Sadov and C. Vafa,``D-Branes and Topological Field
Theories,'' Nucl. Phys. {\bf B463} (1996) 420.}.
Consider in particular the fermions.  Before the twisting
they transform under $SO(3)\times [SO(4)_J\times (U(1)_L\times SU(2)_R)]$
as
$$[s,[(0,\half)_J\otimes (0,\half)]\oplus[(\half,0)_J\otimes (\pm 1,0)]]$$
If we consider the worldvolume of M2 brane to be $\Sigma^g\times S^1$,
this twisting is basically adding a contribution of $U(1)_L^N$
charge to the holonomy.
Since this is related to the $U(1)$ spin connection on the Riemann
surface, due to \scs , it follows that the fermions on
the Riemann surface consist of the following:
\eqn\trfe{\psi\rightarrow [{{\pm 1}\over 2},[(0,\half)_J\otimes (\half)]]\oplus
[2(0),(\half,0)_J\otimes (0)]\oplus [(\pm 1),(\half,0)_J\otimes (0)]}
where the above quantum numbers denote the helicity on the surface,
the $SO(4)_J$ quantum number and the $SU(2)_N^R$ quantum number
respectively (the $U(1)_L^N$ has been combined with the modified spin).
The 8 scalars on M2, which before twisting transform according to
$$[0,[(\half,\half)_J\otimes (0,0)]\oplus[(0,0)_J\otimes (\pm 1,\half)]]$$
after twisting transform according to
\eqn\trsc{[0,[(\half,\half)_J\otimes (0)]]\oplus[{\pm 1\over 2},(0,0)_J\otimes
(\half)]}

In finding the bound states, we need to quantize the zero modes
in the above theory.   Four zero modes for the scalars, corresponding to
$[0,[(\half,\half)_J\otimes (0)]]$ in \trsc\ correspond to the four momenta
in 4+1 dimensions.  The other scalars transforming as
$[{\pm 1\over 2},(0,0)_J\otimes (\half)]$ are paired with fermions with
the same internal quantum numbers.  If there are zero modes for these
directions,
they will correspond to deformations of $\Sigma^g$ in the threefold.
Let us call the corresponding moduli space ${\cal M}$.
Because of the pairing of the fermions with the bosons, this
will give us a supersymmetric sigma model on ${\cal M}$. Note that the
$SO_J(4)$ quantum numbers of these fermions are in $(0,\half)$.  In particular
the $SL(2)$ lefshetz action on the cohomology elements of the Kahler manifold
${\cal M}$ should be identified with the {\it right} $SU(2)$ subgroup
of $SO_J(4)$.
  Finally there
are the zero modes coming from the extra fermions which transform
as scalars and 1-forms on the genus $g$ curve (
$[2(0),(\half,0)_J\otimes (0)]\oplus [(\pm 1),(\half,0)_J\otimes (0)]$).
If the genus $g$ surface is non-degenerate there will be $g+1$ such
zero modes.  Quantizing them gives rise to the representation
whose $SO_J(4)$ quantum numbers are given by
$$[(\half,0)+2(0,0)]^{g+1}$$
Note in particular, that the $SU(2)_L$ content of this
representation is exactly the same as the convenient
basis we discussed before, namely a half hypermultiplet
tensored with $I_g$.

If throughout the moduli space ${\cal M}$ the
Riemann surface remains non-degenerate, the $SO_J(4)$ quantum
numbers of BPS bound states
will be given by
$$[(\half,0)+2(0,0)]^{g+1}\otimes (\sum_{j=0}^{d/2} a_j [(0,j)])$$
where the complex dimension of ${\cal M}$ is denoted by $d$ and
$a_j$ are determined from the $SL(2)$ decomposition.
In particular, the contribution
of this family to $F_g$ will be the same as the contribution
of $(-1)^d\chi ({\cal M})I_g$, as the quantum number of the right $SU(2)$
is counted only as far as its $Z_2$ quantum number $(-1)^{2j_r}$ is concerned.
In other words, in this case we would have
$$\alpha_r=\delta_{g,r}(-1)^d\chi ({\cal M})$$
This result in particular means that an isolated genus
$g$ curve contributes to all $F_k$ for $k\geq g$ according to
\finfor .  This result has also been independently derived recently
from the mathematical
definition of topological strings \ref\panr{R. Pandharipande,
``Hodge Integrals and Degenerate Contributions,'' math.AG/9811140.}.

However quite frequently the moduli space ${\cal M}$ will consist
of loci where the genus $g$ surface has shrunk to a genus $r\leq g$
for some r.  In such cases, even though in principle one can
continue by studying the M-theory description of M2 bound states,
it turns out to be also useful to view it from the type IIA perspective,
by going down on a circle.  This is also useful
for the count of M2 brane bound states.
 As far as the total count of M2 BPS states, we can use the type IIA
computation, as it is radius independent.
However we will have to also recover
the $SU(2)_L$ quantum number of states, which requires additional
care discussed below.

\subsec{The Type IIA Perspective}
The worldvolume theory for a single D2 brane is the same
as that of M2 brane except that instead of 8 scalars, one has 7
scalars and one $U(1)$ gauge field, which is dual to the 8-th
scalar; this is a reflection of duality between scalars and vectors
in 3 dimensions.  As far as constructing bound states and counting
zero modes of wrapped D2 branes, the story is essentially the same as that of
M2 brane, except that now we have only 3 bosonic zero modes, corresponding
to spatial translation in 3 directions.  The effects of the momentum
in the 4-th direction is now replaced by the number of D0-branes
bound to the D2 brane, which in turn is given by the quantum of
flux of the $U(1)$
field strength on the Riemann surface.  Since the choice of the flux
does not affect the degeneracy of the bound states we will recover
the fourth momentum in this way.  There is, however, one extra simplification:
Now the extra $U(1)$ gauge field will have zero modes on a genus $g$
surface $\Sigma$ and in fact will pair up with the unpaired fermionic
zero mode in the M2 brane description.  In other words,
the moduli space ${\cal M}$ of deformations of $\Sigma$
in the Calabi-Yau, is now replaced by $\hat{\cal M}$, which
is the moduli space of deformations of $\Sigma$ together with a choice
of a flat $U(1)$ connection on $\Sigma$.  We thus have a fiber space
$${\hat{\cal M}}\rightarrow {\cal M}$$
with fiber which is generically $T^{2g}$.
 Moreover, the number
of bound states in this formulation is in 1-1 correspondence
with the cohomology of $\hat {\cal M}$. The
spacetime rotation group is $SO(3)$ in this case
and the $SO(3)$ quantum numbers of the D2 brane bound states
can be read off by the Lefshetz $SL(2)$ action
on the cohomologies of $\hat {\cal M}$.  Note that
this $SO(3)$ is the diagonal
$SU(2)$ in $SU(2)_L\times SU(2)_R$ and we would need
to read off this further decomposition from this data.

The basic idea to read off this decomposition of cohomology
is to recall that $SU(2)_L$ refers to the cohomology
in the torus direction (the Jacobian) and the $SU(2)_R$
refers to the cohomology in the base direction.  Let us denote
the base by ${\cal M}$.
The cohomology of ${\hat {\cal M}}$
decomposed taking into account this decomposition to $SU(2)_L\times SU(2)_R$
will consist of:
$$I_g\otimes R_g+I_{g-1}\otimes R_{g-1}+...+I_0 \otimes R_0$$
where $I_g$ is the left representation defined before: $I_k=
[(\half) +2(0)]^k$, and $R_k$ are the right representation which
couple to the left $I_k$.  These will of course depend on the
${\hat {\cal M}}$ and its fibration structure.  The fact that the highest
rank $I$ appearing above is $I_g$ follows from the fact that
highest differential form coming from $T^{2g}$ will have this spin.

The contribution of this configuration to topological string
amplitudes will be given by $\alpha_r$ with $r=0,...,g$ where
$$\alpha_k=\chi(R_k)$$
and $\chi(R_k)$ refers to the sum of the dimension of
$SU(2)_R$ representations in $R_k$ weighted with $(-1)^{2j_R}$.
These already imply the following facts:  If we consider
the Euler characteristic of ${\hat {\cal M}}$ we immediately learn using
 $\chi (I_k\otimes R_k)=\chi(I_k)\otimes \chi (R_k)$
and $\chi (I_k)=\delta_{k,0}$ that:
\eqn\gzer{\alpha_0=(-1)^{{\hat d}}\chi({\hat{\cal M}})}
where ${\hat d}$ is the complex dimension of ${\hat{\cal M}}$.
In other words, the contribution to genus zero of a family of
holomorphic curves is given (up to sign) by the Euler characteristic
of the moduli space of the curve together with the Jacobian
on it.  This fact had already been used in particular to compute the Euler
characteristic
of moduli space of instantons on ${\half K3}$ \ref\mnvw{J.A. Minahan,
D. Nemeschansky, C. Vafa and N.P. Warner,``E-Strings and N=4
Topological Yang-Mills Theories,'' hep-th/9802168.}\
as well as  resolve a puzzle in connection with the counting of black hole
entropy in compactification
of M-theory on Calabi-Yau threefolds \ref\bhva{C. Vafa, ``Black Holes and
Calabi-Yau Threefolds,'' Adv. Theor. Math. Phys. {\bf 2} (1998)
207.}.  For some other applications of this result
see also the recent discussion in
\ref\nekr{N. Nekrasov,``In the Woods of M-theory,'' hep-th/9810168.}.
It is also in line with the observations in \ref\yz{S.T. Yau
and E. Zaslow,``BPS States, String Duality and Nodal Curves
on K3,'' Nucl. Phys. {\bf B471} (1996) 503.}\
that the euler characteristic of a Jacobian variety will localize
on nodal rational curve, as one would expect from a contribution
to the genus zero topological string amplitude $F_0$.

The structure of $R_g$ is also easy to predict:  The fact
that the top cohomology of torus makes invariant sense, implies
that the left spin $[g]$ state will be accompanied by the full
cohomology of the base.  In other words $R_g$ can be obtained
by considering the base manifold ${\cal M}$ and its $SL(2)$ decomposition.
This in particular implies that
$$\alpha_g =(-1)^d\chi ({\cal M})$$
where $d$ is the complex dimension of ${\cal M}$.
In the case of $g=1$ these two ingredients suffice
to find the decomposition of the $SU(2)$ action on
the cohomology of ${\hat {\cal M}}$.  For $g>1$ one would
need to decompose the ${\hat {\cal M}}$ into the cohomology
of the fiber (or a sheaf theory extension of it) times that
of the base.  While this is standard mathematics (known
as Leray spectral sequence \ref\lss{P. Griffiths
and J. Harris, {\it Principles of Algebraic Geometry},
Wiley-Interscience, New York 1978.}) the fact that
the $SU(2)$ action on the fiber continues to make sense
on this decomposition has not been considered previously.
Physically we are predicting that this should be possible
because of the fact that the $SU(2)_L\times SU(2)_R$ quantum
numbers of the cohomology have well defined meanings from
the M-theory perspective on a Calabi-Yau threefold.  It would be interesting
to verify this mathematically.

It is useful to give an example of how things work.  We present
an example which was pointed out to us by Sheldon Katz:  Consider
a ${\bf P}^2$ in a Calabi-Yau threefold.  Consider the contribution
to $F_0$ from primitive degree 3 curves.  This has generically
genus 1.  In this case the ${\cal M}$ is ${\bf P}^9$.  Moreover
${\hat {\cal M}}$ can be studied by viewing it as
follows:  A flat bundle on $T^2$ can be identified
with a point $p$ on the dual torus.  Consider the image
of the point  $p\in {\bf P}^2$.  The moduli space
of elliptic curves passing through that point is ${\bf P}^8$.
  Thus ${\hat {\cal M}}$
in this case can be viewed as a ${\bf P}^8$ bundle over ${\bf P}^2$.
The cohomology of this space is the same as the product of the cohomology
of ${\bf P}^2$ and ${\bf P}^8$.  From what we said before
we thus learn the $SU(2)_L\times SU(2)_R$ decomposition in this case to be
$$[({1\over 2} ,{9\over 2})]+[(0,3)]=[I_1,{9\over 2}]+[(I_0,(3)-2({9\over
2}))]$$
giving us $\alpha_0=27$ and $\alpha_1=-10$.
\subsec{Multiwrapped 2-branes}

As mentioned before, for multiwrapped M2 branes on a Riemann surface
in a Calabi-Yau, it is more convenient to use the D2 brane description,
as that is the {\it definition} of multi-wrapped M2 branes (by taking
the coupling to infinity).  The field content of $N$ D2 branes is
the same as what was noted above, except that now every field is
in addition in the adjoint representation of $U(N)$.
In general the problem of solving for the bound state
will involve the study of the cohomology of the
mixed moduli space of light gauge degrees of freedom
on the D2 brane as well as the scalar zero modes corresponding
to normal deformation of the surface $\Sigma$ in the Calabi-Yau.
This of course will in general be a complicated system to study.

To give a flavor of what new things may happen
when we have multi-wrapped branes, let us consider the case
where the surface $\Sigma$ has no normal deformation in the Calabi-Yau.
In this case the only other zero mode would correspond to the
moduli of the gauge system, which in this case is moduli of
flat $U(N)$ connections on $\Sigma$.  Let us call this space
$K^{g,N}$.  Viewing the gauge group as
$$U(N)={U(1)\times SU(N)\over Z_N}$$
Corresponding to this, the moduli space $K^{g,N}$
splits up to
\eqn\theg{K^{g,N}={T^{2g}\times{\hat K}^{g,N}\over Z_N^{2g}}}
Here $T^{2g}$ is the moduli of flat $U(1)$ connection
on $\Sigma$, which is also known as its Jacobian and
${\hat K}^{g,N}$ is the moduli of flat
$SU(N)$ connections on $\Sigma$.  This consists
of the choice of $2g$ holonomies around the cycles
of the ``opened up'' Riemann surface,
up to conjugation by an $SU(N)$ gauge transformation, satisfying
$$\prod_{i=1}^{g} (g_ih_ig_i^{-1}h_i^{-1})={\rm exp}(2\pi i k/N)$$
giving a space of complex dimension $(g-1)(N^2-1)$.  Here
$k$ is the quantum of $U(1)$ flux through $\Sigma$ which corresponds
to the D0 brane charge.
$Z_N^{2g}$ in \theg\  acts by translations by $1/N$ fractions
along the cycles of $T^{2g}$ at the same time as
changing the corresponding $SU(N)$ holonomy by
an $N$-th root of unity.  If $k$ and $N$ are relatively
prime $K^{g,N}$ is a smooth manifold.  Physically
the non-smoothness when $k$ and $N$ are not relatively
prime can be intepreted as corresponding to the channel
available for a decay of bound states of D0 and D2 brane
at threshold.  To avoid such subtleties it is natural
to work in the case where $k$ and $N$ are relatively prime and expect
the BPS degeneracies to be valid also
when they are not prime.
It is also known that all the cohomological
question of this space is independent of which $k$ one
chooses as long as it is relatively prime with $N$.

Cohomology of the manifold ${\hat K}^{g,N}$ is known
(for a recent discussion
see \ref\mathrc{G. Laumon and M. Rapoport,``The Langlands
Lemma and the Betti Numbers of Stacks of G-bundles on a Curve,''
alg-geom/9503006.}\ and
references therein).  Moreover it
is known that the action of ${Z_N}^{2g}$ is trivial
on the cohomologies.  Thus the cohomologies of $K^{g,N}$
is the same as that of ${\hat K}^{g,N}$ times that of $T^{2g}$.
As far as the $SU(2)_L$ content of the states, it follows
that it is given by
$$I_{g}\otimes I_{{\hat K}^{g,N}}$$
where $I_{{\hat K}^{g,N}}$ denotes the representation of
the corresponding bound
states coming from ${\hat K}^{g,N}$.  Even though apriori
the $SL(2)$ action is the diagonal $SU(2)$ of $SO(4)$, it is natural
to believe that the full gauge degree of freedom is in the $SU(2)_L$
(as is the corresponding $U(1)\subset U(N)$ degrees of freedom)
in which case one can recover the representation $I$ from the
known cohomology formula for ${\hat K}^{g,N}$.

Let us now discuss some properties of ${\hat K}^{g,N}$ for various
$g$. For genus $g=0$ this is an empty space for $N>1$.  In other
words there are no bound states of more than one $D2$ brane around $S^2$.
For $g=1$, ${\hat K}^{g,N}$ is a point for all $N$. In other words
$N$ D2 branes form a single bound state for all $N$.  The
same statement applies to M2 branes, and so we conclude that
$\alpha_1 =1$ and the rest of the $\alpha_i$ for
$i\not= 1$ are zero for all multiwrappings.  Note that this implies,
from \finfor\ that isolated genus 1 curves and their multiwrappings
only contribute to $F_1$.  Thus from a single isolated
elliptic curve with area $A$, and including
its bound state contribution to $D_0$ branes as well
as its multiwrappings we get the $F_1$ contribution
$$-\sum_{n>0}{\rm log}(1-{\rm exp}(-2\pi nA))$$
in accordance with expectations from \bcovi .
Note also that from \finfor\
there is no contribution to $F_g$ for $g>1$, in accordance
with \bcovii.  This result has also been derived recently
using the mathematical definition of topological strings
\panr .

The story is more complicated for genus $g>1$.  The
cohomologies of ${\hat K}^{g,N}$ are known, and so the contribution
of various $N$ to various amplitudes can be obtained using \finfor .
At first sight a puzzle appears:  Consider the contribution of an
isolated genus $g$ curve with area $A$.  As discussed before
it follows from \finfor\ that it contributes in particular to
$F_g$.  The surprising aspect of this is that it will contribute
to $F_g$ also in the form $k^{2g-3}exp(-2\pi k A)$, suggesting
that there is a multicovering of a genus $g$ by a genus $g$ curve.
However this is not possible for $g>1$.  One may have hoped, therefore
that this contribution gets cancelled by the contribution
of $k$ M2 branes bound over that surface.  However it turns out that $
\chi ({\hat K}^{g,N})=0$ and so $I_{{\hat K}^{g,N}}$ will not contain
$I_0$ and so $I_g\otimes I_{{\hat K}^{g,N}}$ will not contain $I_g$
(note $I_k\otimes I_r=I_{k+r}$).
However, it turns out that multicovering of the M2 brane does
have deformations to a higher genus curve.  In other words
multi-covering of an isolated curve is {\it not} isolated and
it can deform to a higher genus surface (said differently
extra scalar zero modes exist if one considers appropriate flat
bundles turned on the D2 branes).  Thus the relevant
moduli space is not just the moduli of flat connection, but it will also
include deformations having to do with a higher genus curve which
does contribute to $F_g$.  Aspects
of this contribution is presently under study \ref\pav{T. Pantev and C.
Vafa, work in progress.}.

We would like to thank  M. Bershadsky, S. Katz, A. Klemm, N.C. Leung,
J. Maldacena,
N. Nekrasov, R. Pandharipande, T. Pantev,
A. Strominger, E. Witten and S.T. Yau for valuable discussions.

The research of R.G. is supported by DOE grant
DE-FG02-91 ER40654 and that of C.V. is supported in part by NSF grant
PHY-98-02709.

\listrefs

\end